# A Low-Power Time-to-Digital Converter for the CMS Endcap Timing Layer (ETL) Upgrade

Wei Zhang, Hanhan Sun, Christopher Edwards, Datao Gong, Xing Huang, Chonghan Liu, Tiankuan Liu, Tiehui Liu, Jamieson Olsen, Quan Sun, Xiangming Sun, Jinyuan Wu, Jingbo Ye, and Li Zhang

*Abstract*—We present the design and test results of a Time-to-Digital-Converter (TDC). The TDC will be a part of the readout ASIC, called ETROC, to read out Low-Gain Avalanche Detectors (LGADs) for the CMS Endcap Timing Layer (ETL) of High-Luminosity LHC upgrade. One of the challenges of the ETROC design is that the TDC is required to consume less than 200 μW for each pixel at the nominal hit occupancy of 1%. To meet the low-power requirement, we use a single delay line for both the Time of Arrival (TOA) and the Time over Threshold (TOT) measurements without delay control. A double-strobe self-calibration scheme is used to compensate for process variation, temperature, and power supply voltage. The TDC is fabricated in a 65 nm CMOS technology. The overall performances of the TDC have been evaluated. The TOA has a bin size of 17.8 ps within its effective dynamic range of 11.6 ns. The effective measurement precision of the TOA is 5.6 ps and 9.9 ps with and without the nonlinearity correction, respectively. The TDC block consumes 97 μW at the hit occupancy of 1%. Over a temperature range from 23 °C to 78 °C and a power supply voltage range from 1.05 V to 1.35 V (the nominal value of 1.20 V), the self-calibrated bin size of the TOA varies within 0.4%. The measured TDC performances meet the requirements except that more tests will be performed in the future to verify that the TDC complies with the radiation-tolerance specifications.

## I. INTRODUCTION

THE High-Luminosity Large Hadron Collider (HL-LHC) operation is currently scheduled to start in 2027. All detectors on the LHC are undergoing upgrades to meet the demands of the HL-LHC operation to achieve the physics goals. One novel detector concept is officially chosen for both the CMS and ATLAS experiments to measure the arrival time of charged particles with a resolution of 30-40 ps per track at the start of the HL-LHC and 60 ps in the worst case at the end of the operation. This new detector, together with the conventional silicon-based tracker, would allow the reconstructed interaction point along the beamline to be well separated in both space and time. In CMS, this is the Minimum-ionizing-particle Timing Detector (MTD) [1]. The MTD has two parts: the Barrel Timing Layer (BTL) and the Endcap Timing Layer (ETL). In the ETL, the Low-Gain Avalanche Detector (LGAD) [2]-[3] is read out by using an Application-Specific Integrated Circuit (ASIC) called the Endcap Timing Read-Out Chip (ETROC) [4]-[5]. A critical design block of the ETROC is a Time-to-Digital Converter (TDC). The TDC performs the precision time measurement of arrival hits against the 40 MHz LHC clock. The TDC obtains the Time of Arrival (TOA) with a less than 30 ps bin size over a 5 ns dynamic range. The TDC also measures the Time Over Threshold (TOT) for time-walk correction [1] with a less than 100 ps bin size over a 10 ns dynamic range. The power consumption of the TDC is required to be less than 200 μW per pixel at the 1% hit occupancy due to the limitation of the cooling system. The power-consumption requirement is most challenging for the TDC design and leads us to study an innovative approach that has never been implemented in ASICs. In this paper, we will present a TDC implemented in a 65 nm CMOS technology. This TDC has achieved a TOA bin size of 17.8 ps over a window of 11.6 ns and a TOT bin size of 35.4 ps over a window of 16.6 ns with a power consumption of 97 μW at the 1% hit occupancy.

## II. DESIGN OF THE TDC

### A. Design approach

The conventional approach of TDC implementation in ASICs is based on a controlled delay line or Delay-Locked Loops (DLLs) [6]-[9]. The delay cells are tuned to match the system clock period. The output taps of the delay cells are routed to a set of registers and the leading edge of the input hit signal captures the delay pattern. An advantage of this approach is that the propagation delay of each delay cell is well known for the cost of extra power to control the delay cells. If the speed of the delay cells is not high enough, a Vernier architecture can be used to improve the precision. However, it is hard for a TDC implemented in this approach to meet the low-power-consumption requirement of the ETROC.

Our TDC design is based on a simple untuned delay-line approach originally developed in the Field Programmable Gates

Wei Zhang, Hanhan Sun, Xing Huang, Li Zhang are with Southern Methodist University, Dallas, TX 75275 USA and also with Central China Normal University Wuhan, Hubei 430079, China. (corresponding authors: Datao Gong and Jinyuan Wu).

Datao Gong, Chonghan Liu, Tiankuan Liu, and Jingbo Ye are with Southern Methodist University, Dallas, TX 75275 USA (e-mail: dgong@smu.edu).

Christopher Edwards, Tiehui Liu, Jamieson Olsen, Quan Sun, and Jinyuan Wu are with Fermi National Accelerator Laboratory, *Batavia*, IL 60510 USA (e-mail: jywu168@fnal.gov).

Xiangming Sun is with Central China Normal University Wuhan, Hubei 430079, China.

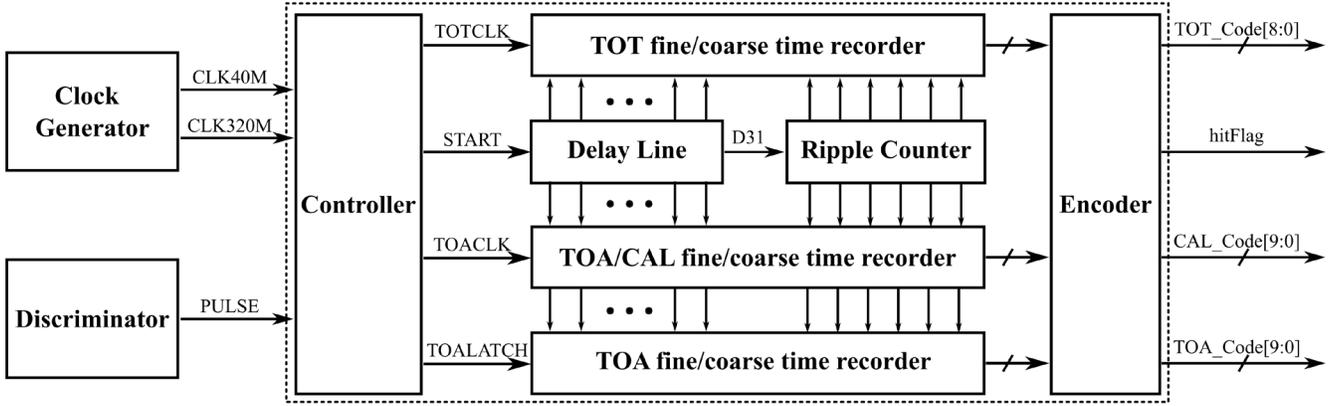

Fig. 1. Block Diagram of the TDC.

Arrays (FPGAs) [10]-[12]. In this approach, the input hit signal propagates in the delay line and a register array takes snapshots at the leading edge of the system clock. The position of the signal transition in the delay line is encoded into the arrival time relative to the clock. Due to the simplicity of the untuned delay line, the delay cell is fast and power-efficient. We use one single delay line to measure both the TOA and the TOT.

The propagation delay of each untuned delay cell is sensitive to process variation, temperature, and power supply voltage. To compensate for these variations, a double-strobe self-calibration approach successfully demonstrated in FPGAs is used to record the timestamp twice. For each input hit signal, the extra measurement calibrates the propagation delay and improves measurement precision.

When the propagation of the whole delay line is shorter than the required range of the TOA and the TOT, after the input hit reaches the end of the delay line, it is fed back to the beginning of the delay line and oscillates in the delay line. The oscillation stops after both the TOA and the TOT measurements are complete. A counter is incremented to count the oscillation. Accurate TOA and TOT measurements are then calculated by combining the coarse counter values with a snapshot of the delay line (the fine values).

*B. Design concept*

Fig. 1 shows the block diagram of the TDC. The input hit signal (PULSE) of the TDC is the discriminator output after the preamplifier signal. The TDC measures the TOA and the TOT of the PULSE signal. The TDC needs a pair of external clock signals, one at 40 MHz (CLK40M) and the other one at 320 MHz (CLK320M), from an on-chip clock generator. Based on the signals CLK40M, Clk320M, and PULSE, a controller generates a START signal, a TOA clock (TOACLK), a TOT clock (TOTCLK), and a TOA latch signal (TOALATCH). The time measurement is implemented with a delay line for fine time measurement and a ripple counter for coarse time measurement. Three recorders, each implemented in a D-Flip-Flop (DFF) chain, snapshot the fine and coarse times of the TOA, the TOT, and the calibration, respectively. The data are encoded in a TDC encoder. The hit flag indicates if there are valid data in a 40 MHz clock cycle.

Fig. 2 displays the timing diagram of the major signals used in the TDC. CLK40M is a copy of the 40 MHz LHC clock. CLK320M has two successive pulses with a known phase difference every 25 ns. The phase difference is programmable with a default value of 3.125 ns. The leading edge of the first pulse of CLK320M is aligned with the rising edge of CLK40M. The START signal is asserted after the leading edge of the PULSE signal and reset after the leading edge of the second pulse of the CLK320M signal. The assertion of START enables the oscillation of the delay line and the de-assertion of START disables the oscillation. The TOACLK signal has two consecutive pulses with the same phase difference as CLK320M. The first pulse is asserted after the leading edge of the CLK40M signal and the second pulse is asserted after the second pulse of CLK320M. The TOALATCH signal duplicates the first pulse of the TOACLK signal. The TOACLK signal snapshots two timestamps in the TOA/CAL recorder. The first timestamp stores the TOA information, while the second one keeps the calibration information. The TOALATCH signal transfers the first timestamp immediately to the TOA recorder. The second timestamp is stored in the TOA/CAL recorder for in-situ calibration. The leading edge of the TOTCLK is aligned with the trailing edge of the signal PULSE and is used to catch the TOT state in the delay line. To minimize the time of the DFFs in the transparent mode and save power, both the TOACLK and TOTCLK signals are active low for about 400 ps.

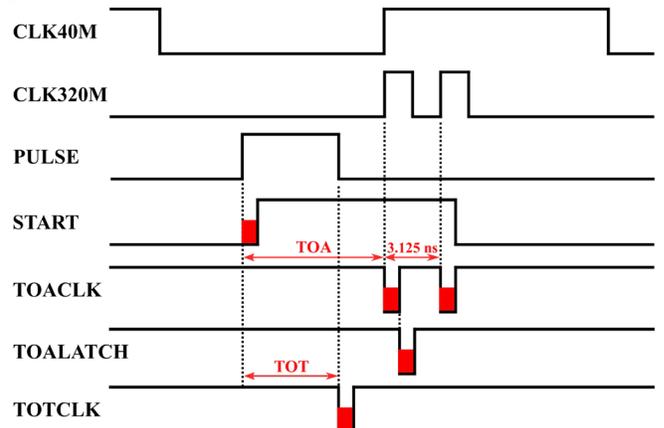

Fig. 2. Timing diagram of the major signals.

*C. Fine time measurement*

Fig. 3 is the schematic of the TDC fine time measurement circuit. The delay line has 63 NAND gates and forms an oscillator. All gates are configured as inverters except that the first one is controlled by the START signal. At each rising edge of the TOACLK or TOTCLK signals, a DFF chain takes a snapshot of the delay line for the fine time information. The resolution requirement of the TOT is more relaxed than that of the TOA, so the TOA DFF chain attaches to every tap (D0, D1, D2, …, D62) of the delay line, while the TOT DFF chain only connects to every other tap (D0, D2, D4, …, D62).

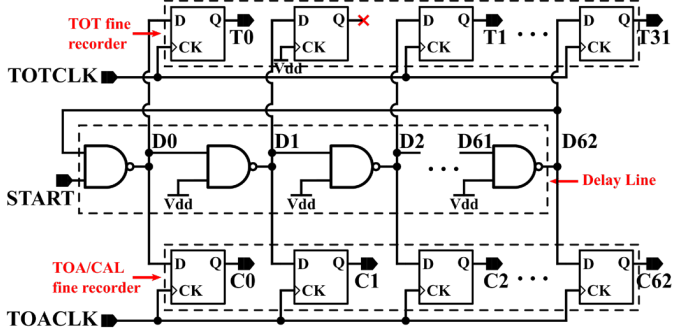

Fig. 3. Schematic of the fine time measurement circuit.

*D. Coarse time measurement*

The coarse time is derived from a 3-bit ripple counter, which records the number of oscillating turns. Fig. 4 is the schematic of the TDC coarse time measurement. The clock of the ripple counter comes from the 31$^{st}$ tap (D31) of the delay line. The signal D31 is buffered by a latch, which has the same load as a DFF. The coarse time measurement is implemented in a 3-bit ripple counter and two DFF chains. The two DFF chains snapshot the timestamps of the coarse time information for the TOA and the TOT, respectively.

Since the START signal is asynchronous with the CLK40M and CLK320M signal, the TOA/TOT recorders may violate the setup/hold time requirements of the DFFs. In this case, the metastability phenomenon may occur and the outputs of the DFFs may be unstable. To resolve the metastability problem, be noted that the metastability also exists in the fine time recorders. However, at any specific moment, only one DFF in each recorder may have a metastable state. After its unstable output is considered in the encoder design, metastability affects only the Least Significant Bits (LSBs) of the fine time measurement.

*E. Calibration circuit*

The schematic of the calibration and the TOA recorders is shown in Fig. 5. As mentioned before, the TOACLK signal has two consecutive pulses with a known phase difference, which carries the calibration information. The fine and coarse time measurements are snapshotted by two DFF chains, as shown in Fig. 3 and 4. The DFF chains that snapshot the TOA time information are triggered twice. The first timestamp, which represents the TOA information, is stored in the TOA/CAL DFF chain, while the second timestamp, which stands for the calibration information, is kept in the calibration TOA DFF chain.

*F. Prototype chip*

The TDC block includes a few supporting functional blocks that are used for test purposes. Electrical transmitter/receiver (eTx/eRx) ports transmit/receive differential signals. A clock divider divides an input clock of 1.28 GHz to generate the internal clocks at 40 MHz and 320 MHz. A Diagnostic-Mode Read-Out (DMRO) circuit scrambles the 30-bit TDC data (including the TOA code, the TOT code, the CAL code, and the hit flag), adds a 2-bit frame header (2'b10), and serializes the 32-bit parallel data to a 1.28 Gigabit per second (Gbps) serial stream in each 40 MHz clock cycle. An I$^2$C module is used for slow control. A Gated Ring Oscillator (GRO) is a copy of the delay line and can be used to calibrate the oscillation frequency of the delay line.

Fig. 6(a) shows the layout of the TDC block. The dimension of the TDC block is 0.700 mm × 1.900 mm. The TDC core, excluding input/output pads and supporting circuits, occupies 0.169 mm × 0.468 mm.

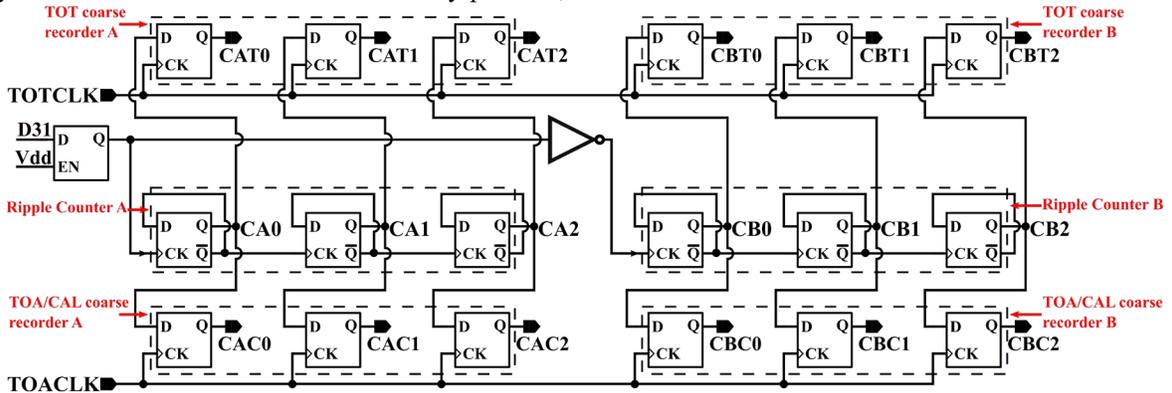

Fig. 4. Schematic of the coarse time measurement.

the counter and the recorders are duplicated, whereas the clock of the duplicated counter is inverse. If the metastability occurs in one recorder groups, the outputs of the other recorder group are stable. The fine time information is used to distinguish which counter recorders are more stable than the other. It should

The prototype of ETROC named ETROC1 is designed and fabricated in a 65 nm CMOS technology [13]. Besides a dedicated TDC block, the ETROC1 contains a 5 × 5 array of pixels. Each pixel consists of a preamplifier, a discriminator, and a standalone TDC block along with digital readout circuitry. Fig.

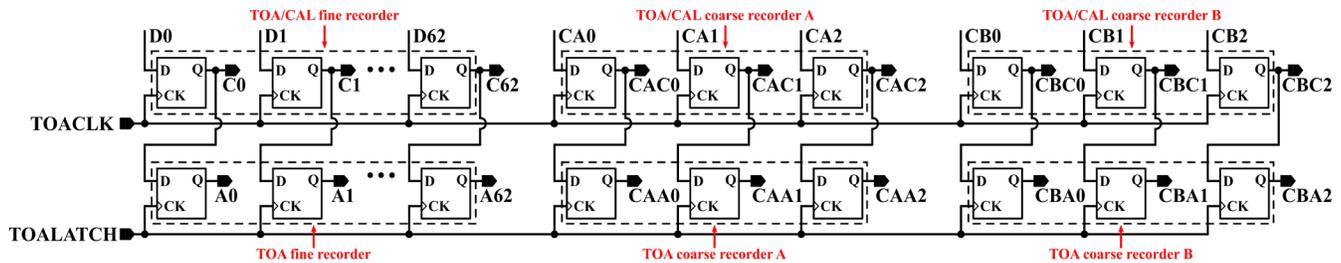

Fig. 5. Schematic of the calibration and the TOA recorders.

6(b-c) shows the photographs of the TDC block and the ETROC1 wire bonded to a test board. The TDC block is located on the upper right corner of the ETROC1.

The test results of the standalone TDC will be reported in this paper. The design and the test results of the ETROC1 are beyond the scope of this paper and will be reported somewhere else in the future.

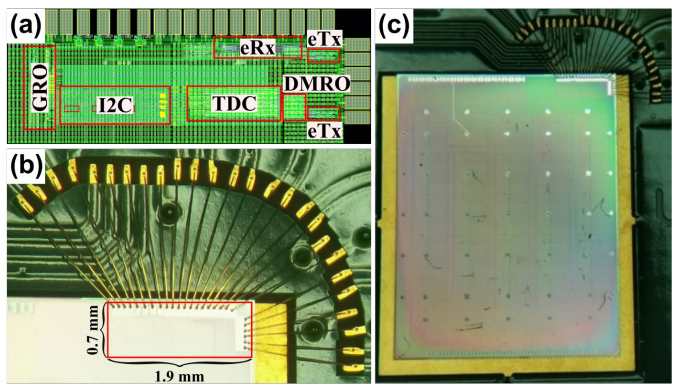

Fig. 6. Layout (a), photograph of the TDC block (b), and photograph of ETROC1 (c).

## III. TEST RESULTS OF THE TDC

### A. Test Setup

The block diagram of the test setup is shown in Fig. 7. In our test, a pattern generator (Centellax Model PCB 12500) provides a 1.28 GHz clock and a 40 MHz clock. The TDC uses the 1.28 GHz clock to divide into the internal 40 MHz and 320 MHz clocks. Based on the 40 MHz clock of the pattern generator, a clock builder (Silicon Laboratory Model SI5338EVB) produces a clock with the required frequency and a pulse generator supplies the input pulse of the TDC with the required width. The jitter of the input pulse is about 2 ps (RMS), significantly less than the TOA/TOT precision. The pulse requirements are different for dedicated TOA and TOT measurements. For the dedicated TOA measurement, the pulse frequency emulates the hit occupancy of the ETROC in the CMS application. We choose a frequency that is correlated to but slightly higher than a sub-frequency of the system clock. For example, for the 1% hit occupancy, the frequency of the input pulse is 400.01 kHz. The small frequency difference makes the arrival time of the input pulse chase the system clock of 40 MHz constantly with a step of 0.625 ps so that the entire measurement range is evenly covered [14]. The pulse width is 6.25 ns, controlled in the pulse generator (Tektronix Model DPG5274). For the dedicated TOT measurement, the pulse frequency is fixed at 40 MHz. The pulse width is adjusted in the pulse generator (Hewlett Packard Model 8133A). The pulse width increases 2 ps from 0.4 ns after each data acquisition until the width reaches the maximum 10.2 ns set by the pulse generator and starts over. For both the dedicated TOA and TOT measurements, the output data of the TDC at 1.28 Gbps are buffered in an FPGA (Xilinx Model KC705 Development Kit) and transferred to a personal computer (PC) through an Ethernet cable.

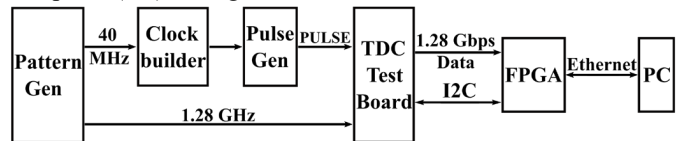

Fig. 7. Block diagram of the TDC test setup

Fig. 8 shows the photographs of the test setup. The standalone TDC block of the ETROC1 chip is wire bonded on the test board, allowing direct access to the signals for the standalone TDC block.

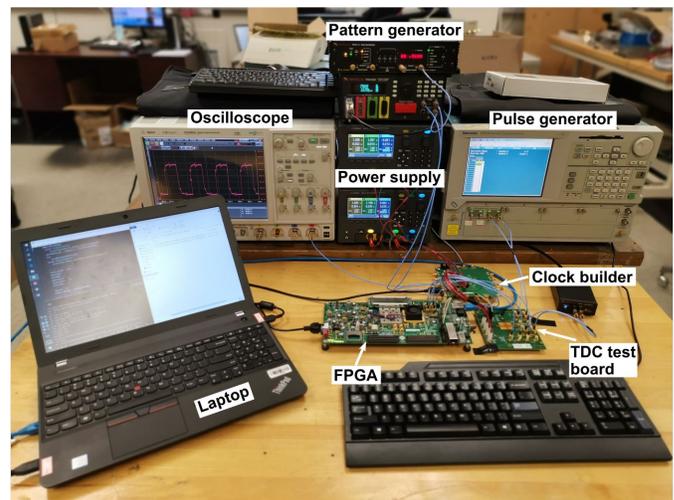

Fig. 8. Photographs of the TDC test setup.

### B. Resolution

In the TOA (TOT) measurement, the arrival time (width) of the input pulse increases a fixed value after each clock cycle (data acquisition). The TOA/TOT times are calculated based on the index of the acquired data. If we plot the measured TOA/TOT codes with all TOA/TOT times, we obtain a stair-shape curve, namely the transfer function.

Fig. 9 shows the transfer functions of the TOA and the TOT. The details of each transfer function at a typical TOA (TOT) code of 56 (150) are zoomed in and displayed in the lower right corner. The effective dynamic range of the TOA is about 11.6 ns, significantly larger than the required range of 5 ns. The large dynamic window of the TOA can be used to study long-lived particles in the HL-LHC. The effective dynamic range of the TOT is measured separately to be 16.6 ns, larger than the range, which is limited by the pulse generator during the test, shown in the figure. The effective dynamic range of the TOT is larger than the required dynamic range of 10 ns.

A minimum-square linear fitting of the measured data is also shown in the figure. The reciprocal of the linear fitting slope is the quantization resolution or the average bin size of the measurement. The bin size represents the propagation delay of each delay cell. The resolution of the TOA is estimated to be 17.8 ps. The bin size of the TOT is 35.4 ps, around twice the TOA bin size as expected.

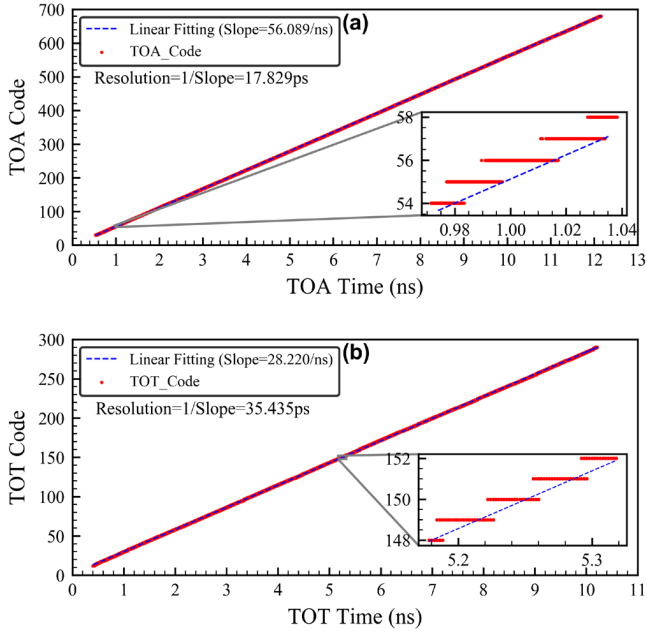

Fig. 9. Measured transfer functions of the TOA and the TOT measurements.

### C. Nonlinearity

Nonlinearity is the deviation of the transfer function from the ideal linear relationship. Nonlinearity is usually divided into Integral Non-Linearity (INL) and Differential Non-Linearity (DNL).

The INL reflects the deviation of each stair center in the transfer function from the anticipated time. Fig. 10 shows the INLs of the TOA and the TOT. Based on our measurements, the INL of the TOA (TOT) is less than ±1.0 LSB (TOA) and ±1.3 LSB (TOT). The INL performances meet the ETROC specifications.

The DNL reflects the uneven stair width of the transfer function. In our measurement, the arrival time of the pulse and the pulse width are uniformly distributed in the effective windows. However, due to the existence of the DNL, the distribution of the TOA and the TOT codes are not uniform. The count of each code is proportional to its bin size. The DNL is estimated as follows:

$$DNL(i) = \frac{count(i)}{average\ count} - 1 \qquad (1),$$

where $DNL(i)$ is the DNL of the $i$th code of the TOA or TOT, $count(i)$ is the count of the $i$th code, and the *average count* is the average count of all codes. Fig. 11 shows the DNLs of the TOA and the TOT. Based on our measurements, the DNL is less than ±0.5 LSB and ±0.8 LSB for the TOA and the TOT, respectively. The DNL performances of the TOA/TOT meet the ETROC specifications.

For the TOA measurement, the DNL represents the nonuniform propagation delay of the NAND delay cells. It can be observed that the delay of the even cells is slightly different from that of the odd cells. This even-odd effect reflects the fact that the propagation delay from the rising edge to the falling edge is different from the one from the falling edge to the rising edge. The DNL of the TOA has a period of 126, combining the delay line of 63 NAND gates with the even-odd effects.

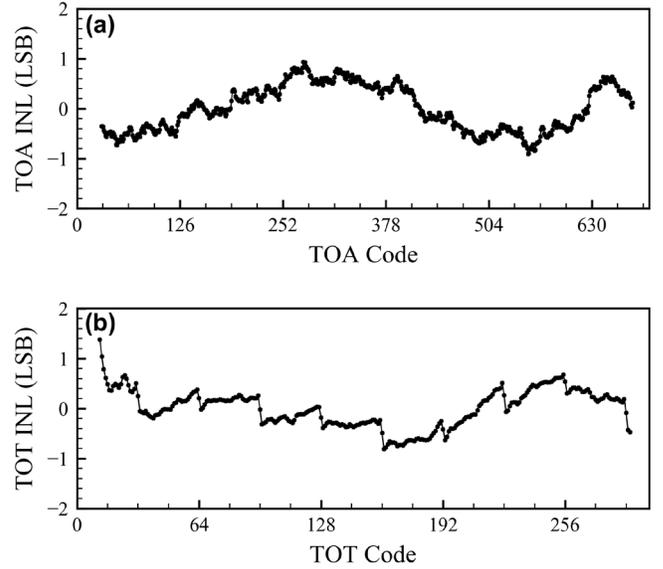

Fig. 10. INL of the TOA (a) and the TOT (b) measurements.

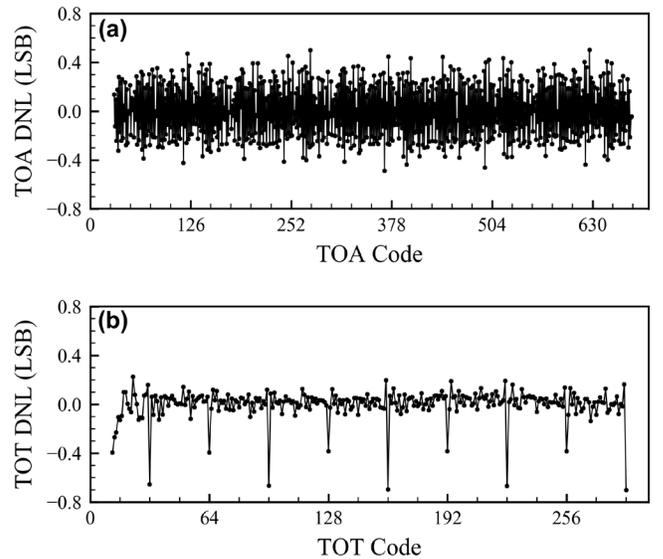

Fig. 11. DNL of the TOA (a) and the TOT (b) measurements.

The same periodicity exists in the TOT DNL. Since the TOT

DFF chain only connects to every other tap of the delay cells, the TOT code from 1 to 31 and from 33 to 64 is double the propagation delay of a NAND gate, whereas the bin size of the TOT at 32 or 64 is the propagation delay of the first NAND gate (in the TDC implementation, the TOT code has an offset of 1). That's why the values of the DNL at 32, 64, 96, … are less than the other values. The periodicity of the TOT DNL is also reflected in the TOT INL figure, where there is a jump after every 32 codes.

### D. Effective measurement precision

The residuals of the linear fitting of the TOA/TOT transfer function represents the deviation of the measurements from the anticipated values. The histograms of the linear fitting residuals at the TOA (TOT) of 56 (150), typical values of the TOA/TOT codes in the ETL application, are shown in Fig. 12(a).

Fig. 12(b) displays the histograms of the combined fitting residuals of all TOA/TOT codes. The standard deviation of the combined residuals is about 9.9 ps (16.7 ps), which represents the effective measurement precision in a large dynamic range without nonlinearity correction.

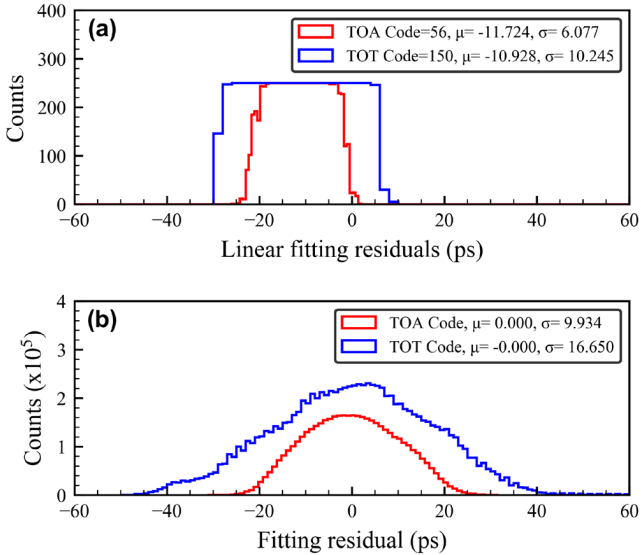

Fig. 12. Histograms of the linear fitting residuals at a typical TOA/TOT code (a) and all TOA/TOT codes (b).

When we operate the TDC in a large dynamic range, the effective measurement precision with nonlinearity correction is calculated as the weighted average [14] of the standard deviation of all codes as follows:

$$\sigma = \sqrt{\frac{\sum_i w(i)\sigma^2(i)}{\sum_i w(i)}} \quad (2),$$

where $\sigma$ is the weighted average of all codes, $\sigma(i)$ is the standard deviation of the $i$th code, and $w(i)$ is the bin size of the $i$th code. The effective measurement precision of the TOA (TOT) measurement is estimated to be about 5.6 ps (10.4 ps). Based on the average bin size, the quantization noise of the TOA (TOT) is estimated to be 5.2 ps (10.2 ps). The quantization noise dominates the effective measurement precision with nonlinearity correction. When we operate the TDC in a small window (e.g., the TOA measurement in the CMS ETL application), the effective measurement precision without the nonlinearity correction is close to the effective measurement precision with nonlinearity correction.

### E. Power Consumption

We have measured the power consumption of the TDC functional block for various operating conditions. At the hit occupancies of 1%, 2%, 5%, and 10%, the total power consumptions of the TDC block are 97, 122, 195, 318 μW, respectively. The power consumption of 97 μW at the nominal hit occupancy of 1% is less than half of the specification.

### F. Self-calibration

The delay cells used in the TDC block are plain CMOS NAND gates and their propagation delays change as the process variations, power supply voltage, and temperature. However, using two strobes separated with a known time interval, the propagation delays are measured constantly. Therefore, the variation of the propagation delays of the delay cells can be self-calibrated during the offline analysis process.

Fig. 13 shows the histogram of the measured calibration codes in the TOA test. The measured calibration codes fall into four consecutive bins. The mean value (173.5) of the calibration codes is consistent with the average propagation delay (17.8 ps) of each delay cells within 1%.

We have measured the bin sizes of the TDC at various power supplies voltages (VDDs) from 1.05 V to 1.35 V and room temperature 23 °C. We have also examined the bin sizes of the TDC at various temperatures from 23 °C to 77 °C and the nominal power supply voltage of 1.20 V. The measured bin sizes of the TOA and the calibration versus VDDs and temperatures are shown in Fig.14.

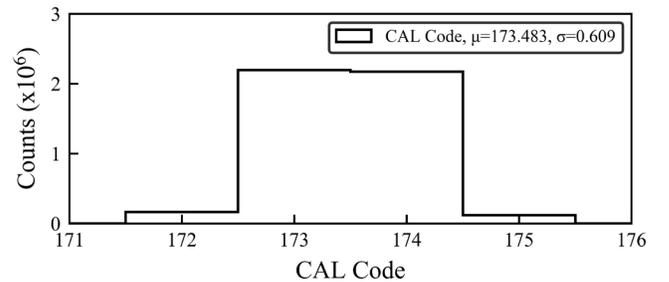

Fig. 13. Histogram of the measured calibration code in the TOA measurement.

As can be seen in the figure, the bin sizes of the TOA and the calibration decrease with the power supply voltage and increase with temperature, consistent with our expectations. The calibration bin size is slightly larger than the TOA bin size. The bin size difference between the TOA and the calibration becomes more significant when the power supply voltage goes lower. This trend implies that the bin size difference results from the power supply voltage decrease. This hypothesis is confirmed in the simulation. In our post-layout simulation, the internal power supply voltage of the delay line decreases slightly after the other parts of the TDC than the delay line start working. The power supply voltage of the delay line could be separated from those of the rest parts of the TDC if we desire to improve the bin size uniformity.

The worst-case difference between the TOA bin size and the calibration bin size is 1.5% at 1.05 V and the room temperature. Such a bin size difference between the TOA and the calibration can be tolerated even if it is not calibrated. However, the difference can be easily self-calibrated offline.

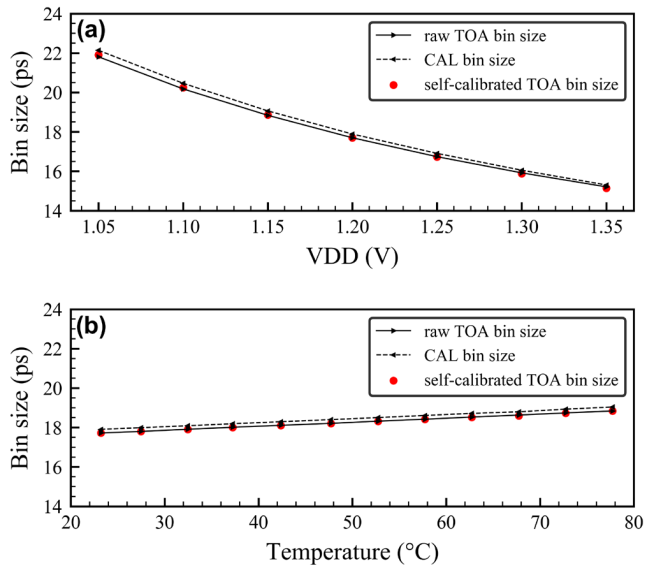

Fig. 14. Bin sized of raw TOA, calibration, and self-calibrated TOA at different power-supply voltages (a) and temperatures (b).

At room temperature, from 1.05 V to 1.35 V, the bin size of the TOA varies 23%, relative to the bin size at the nominal power supply voltage. Similarly, at the nominal power supply voltage, from 23 °C to 77 °C, the bin size of the TOA varies 6.3%, relative to the bin size at room temperature. Such the bin size variation makes the offline self-calibration critically important to achieve the required measurement precision.

The self-calibration has been demonstrated to work well in the test. The bin size of the TOA can be self-calibrated per the following equation:

$$BIN_{TOA\ cal}(x) = \frac{BIN_{TOA\ meas}(x_0)}{BIN_{CAL}(x_0)} \cdot BIN_{CAL}(x) \quad (3),$$

where $BIN_{TOA\ cal}$ is self-calibrated TOA bin size, $BIN_{TOA\ meas}$ is the raw TOA bin size, $BIN_{CAL}$ is the calibration bin size, x represents the power supply voltage VDD or the temperature, and $x_0$ is the nominal power supply voltage of 1.20 V or room temperature. The self-calibrated TOA bin size is also shown in the figure. Comparing to the raw TOA bin size, the self-calibrated TOA bin sizes vary within 0.43% and 0.16% in the whole range of the power supply voltage and temperature, respectively.

*G. Radiation Tolerance*

The TDC is required to tolerate a Total Ionizing Dose (TID) of 1 MGy. Because of no access to the irradiation facility (more information available at https://espace.cern.ch/project-xrayese/_layouts/15/start.aspx#/) at European Organization for Nuclear Research (CERN) during the Covid-19 pandemic, we preliminarily evaluate the radiation tolerance of the TDC with a low-dose-rate facility at Southern Methodist University. More tests will be performed in the future to verify that the TDC meets the radiation tolerance specifications.

The TDC was exposed in an x-ray irradiator (Precision x-rays, Inc., Model iR160) with the maximum energy of 160 keV for about 7 hours. Fig. 15 shows a photograph of the test setup. The TID reached 23.4 kGy with a dose rate of 0.93 Gy/s.

During the irradiation test, the power supply current was monitored every 5 seconds. Fig. 16(a) shows the change of the power supply current related to its initial value before the test. As can be seen in the figure, the power-supply current decreases within 5%.

During the test, a pulse with a width of 6.25 ns and a frequency of 80 MHz was input to the TDC. The TOA, the TOT, and the calibration codes of 500 words each were acquired every 5 seconds. The mean value, the minimum value, and the maximum value of each code in each acquisition were calculated. The change of the TOA, the TOT, and the calibration are shown in Fig. 16(b-d). As can be seen in the figure, the TOA, the TOT, and the calibration codes decreased within 1 LSB and overwhelmed in quantization noise.

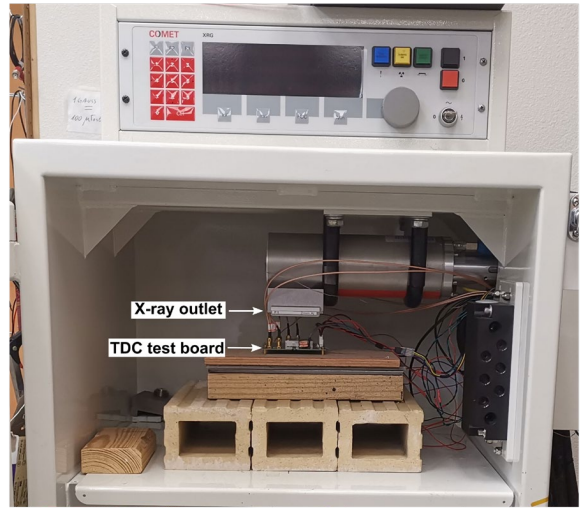

Fig. 15. Photograph of the irradiation test setup.

All performances of the TDC were measured before and after the test and compared. The bin sizes of the TOA, the TOT, and calibration decreased consistently by 0.67%, 1.2%, and 0.68%, respectively. It should be noted that the change of the TOA/TOT codes after the test is in the inverse direction of the change observed during irradiation. Therefore, the changes of the TOA/TOT bin sizes after the test are not induced by radiation and more likely due to the test environments such as the temperature or the power supply voltage. No significant degradation was observed in the precision, the INL, and the DNL of the TOA/TOT.

## IV. CONCLUSION

We have successfully designed and tested a delay-line based TDC in a commercial 65 nm CMOS technology. The TOA has a bin size of 17.8 ps within its effective dynamic range of 11.6 ns. The Differential Non-Linearity (DNL) and the Integral Non-Linearity (INL) of the TOA are less than ±0.5 LSB and ±1.0 LSB, respectively. The effective measurement precision of the TOA is 5.6 ps and 9.9 ps with and without the nonlinearity correction, respectively. The TOT has a bin size of 35.4 ps. The dynamic range of the TOT is measured to be 16.6 ns. The DNL and INL of the TOT are ±0.8 LSB and ±1.3 LSB, respectively. The effective measurement precision of the TOT is 10.4 ps and 16.7 ps with and without the nonlinearity correction, respectively. The actual TDC block consumes 97 µW at the hit occupancy of 1%. The self-calibration tests demonstrate that over a temperature range from 23 °C to 78 °C and a power supply voltage range from 1.05 V to 1.35 V, the self-calibrated bin size varies within 0.4%. The TDC survives the TID of 23.4 kGy. The measured TDC performances meet the requirement for the ETROC for the CMS ETL project except for the radiation tolerance. More tests will be performed in the future to verify that the TDC complies with the radiation tolerance specifications.

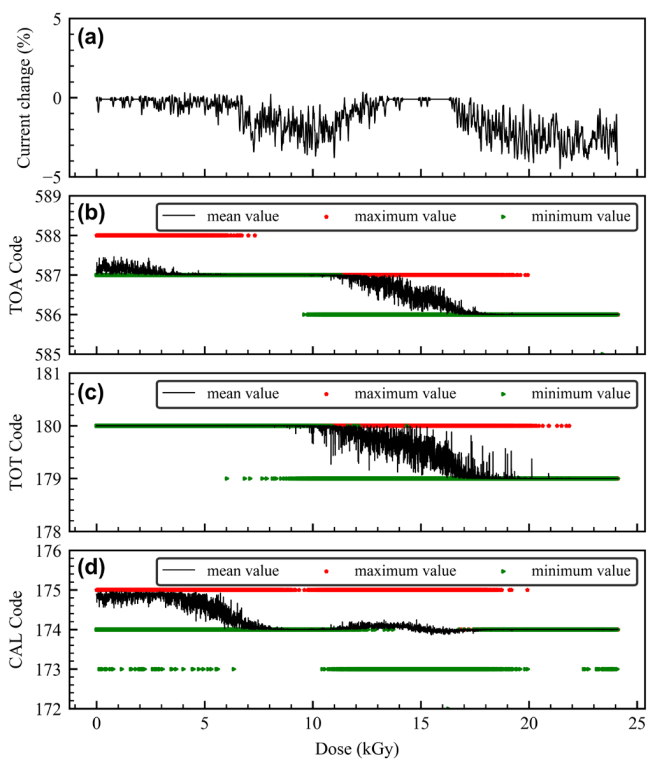

Fig. 16. Relative change observed during the irradiation test of the power supply current (a), change of the TOA code (b), the TOT code (b), and the calibration code (d).

ACKNOWLEDGMENT

We are grateful to Dr. Paulo Moreira, Dr. Rui Francisco, and Dr. Szymon Kulis of CERN for sharing the ELT library in lpGBT and the generic I2C module for the TDC configuration.

REFERENCES

[1] The CMS collaboration, "A MIP timing detector for the CMS Phase-2 upgrade technical design report," *CERN-LHCC-2019-003, CMS-TDR-020*, Sep. 2019. Available at https://cds.cern.ch/record/2667167?ln=en.
[2] G. Pellegrini et al., "Technology developments and first measurements of Low Gain Avalanche Detectors (LGAD) for high energy physics applications," *Nucl. Instrum. Meth.* A, vol. 765, pp. 12-16, Nov. 2014, DOI:10.1016/j.nima.2014.06.008.
[3] N. Cartiglia et al., "Design optimization of ultra-fast silicon detectors," *Nucl. Instrum. Meth.* A, vol. 796, pp. 141-148, Oct. 2015, DOI:10.1016/j.nima.2015.04.025.
[4] Tiehui Liu et al., "The ETROC Project: ASIC development for CMS Endcap Timing Layer (ETL) upgrade," Topical Workshop on Electronics for Particle Physics (TWEPP), Santiago de Compostela, Spain, Sep. 2019. Available at https://indico.cern.ch/event/799025/contributions/3486223/.
[5] Quan Sun, Sunil M. Dogra, Christopher Edwards, Datao Gong, Lindsey Gray, Xing Huang *et al.*, "The Analog Front-end for the LGAD Based Precision Timing Application in CMS ETL," *22nd Virtual IEEE Real Time Conference*, Oct. 2020. Available at: https://indico.cern.ch/event/737461/contributions/4013066/attachments/2120645/3570531/P_A-02_226.pdf.
[6] M. Mota and J. Christiansen, "A high-resolution time interpolator based on a delay locked loop and an RC delay line," *IEEE J. Solid-State Circuits, vol.* 34, pp. 1360-1366, Oct. 1999, DOI: 10.1109/4.792603.
[7] P. Dudek, S. Szczepanski, and J. Hatfield, "A high resolution CMOS time-to-digital converter utilizing a Vernier delay line," *IEEE J. Solid State Circuits,* vol. 35, pp. 240–247, Feb. 2000, DOI: 10.1109/4.823449.
[8] N. Seguin-Moreau et al., "ALTIROC1, a 25 pico-second time resolution ASIC for the ATLAS High Granularity Timing Detector (HGTD)," *PoS vol. 370-TWEPP 2019,* p. 042, Apr. 2020, DOI: 10.22323/1.370.0042.
[9] Paweł Kwiatkowski and Ryszard Szplet, "Efficient Implementation of Multiple Time Coding Lines-Based TDC in an FPGA Device," *IEEE Trans. Instrum. Meas.*, vol. 69, pp. 7353-7364, Oct. 2020, DOI: 10.1109/TIM.2020.2984929.
[10] Jinyuan Wu and Zonghan Shi, "The 10-ps wave union TDC: Improving FPGA TDC resolution beyond its cell delay," *2008 IEEE Nuclear Science Symposium Conference Record*, Dresden, Germany, pp. 3440-3446, Oct. 2008, DOI: 10.1109/NSSMIC.2008.4775079.
[11] J. Wu, "Several Key Issues on Implementing Delay Line Based TDCs Using FPGAs," *IEEE Trans. Nucl. Sci.*, vol. 57, pp. 1543-1548, Jun. 2010, DOI: 10.1109/TNS.2010.2045901.
[12] J.Wu, Y. Shi, and D. Zhu, "A low-power Wave Union TDC implemented in FPGA," *J. Instrum.* vol. 7, p. C01021, Jan. 2012, DOI: 10.1088/1748-0221/7/01/C01021.
[13] L. Jara Casas et al., "Characterization of radiation effects in 65 nm digital circuits with the DRAD digital radiation test chip," *J. Instrum.,* vol. 12, p. C02039, Feb. 2017, DOI: 10.1088/1748-0221/12/02/C02039.
[14] Jinyuan Wu, "Uneven Bin Width Digitization and a Timing Calibration Method Using Cascaded PLL," *2014 19th IEEE-NPSS Real Time Conference*, Nara, Japan, May 2014, DOI: 10.1109/RTC.2014.7097534.